\documentclass[sigconf]{acmart}
\usepackage[english]{babel}

\usepackage{bm}
\usepackage{amsmath}
\usepackage{multirow}
\usepackage{booktabs}
\usepackage{adjustbox}
\usepackage{float}
\usepackage{array}
\usepackage[justification=justified,skip=2pt]{caption}

\newcommand{\PreserveBackslash}[1]{\let\temp=\\#1\let\\=\temp}
\newcolumntype{C}[1]{>{\PreserveBackslash\centering}p{#1}}
\newcolumntype{R}[1]{>{\PreserveBackslash\raggedleft}p{#1}}
\newcolumntype{L}[1]{>{\PreserveBackslash\raggedright}p{#1}}

\renewcommand{\arraystretch}{1}
\newcommand{\etal}{\textit{et al}.}

\AtBeginDocument{%
  \providecommand\BibTeX{{%
    \normalfont B\kern-0.5em{\scshape i\kern-0.25em b}\kern-0.8em\TeX}}}

\begin{document}

\title{NGAT4Rec: Neighbor-Aware Graph Attention Network For Recommendation}

\author{Jinbo Song}
\affiliation{\institution{Institute of Computing Technology, Chinese Academy of Sciences}}
\email{songjinbo18s@ict.ac.cn}

\author{Chao Chang}
\affiliation{\institution{Beijing Kuaishou Technology Co., Ltd.}}
\email{changchao@kuaishou.com}

\author{Fei Sun}
\affiliation{\institution{Alibaba Group}}
\email{ofey.sunfei@gmail.com}

\author{Xinbo Song}
\affiliation{\institution{Institute of Computing Technology, Chinese Academy of Sciences}}
\email{songxb@ict.ac.cn}

\author{Peng Jiang}
\affiliation{\institution{Beijing Kuaishou Technology Co., Ltd.}}
\email{jiangpeng@kuaishou.com}

\renewcommand{\shortauthors}{Song \etal}

\begin{abstract}
Learning informative representations (aka. embeddings) of users and items is the core of modern recommender systems. Previous works exploit user-item relationships of one-hop neighbors in the user-item interaction graph to improve the quality of representation. Recently, the research of Graph Neural Network (GNN) for recommendation considers the implicit collaborative information of multi-hop neighbors to enrich the representation. However, most works of GNN for recommendation systems do not consider the relational information which implies the expression differences of different neighbors in the neighborhood explicitly. The influence of each neighboring item to the representation of the user’s preference can be represented by the correlation between the item and neighboring items of the user. Symmetrically, for a given item, the correlation between one neighboring user and neighboring users can reflect the strength of the signal about the item’s characteristic. To modelling the implicit correlations of neighbors in graph embedding aggregating, we propose a Neighbor-Aware Graph Attention Network for recommendation task, termed NGAT4Rec. It employs a novel neighbor-aware graph attention layer that assigns different neighbor-aware attention coefficients to different neighbors of a given node by computing the attention among these neighbors pairwisely. Then NGAT4Rec aggregates the embeddings of neighbors according to the corresponding neighbor-aware attention coefficients to generate next layer embedding for every node. Furthermore, we combine more neighbor-aware graph attention layer to gather the influential signals from multi-hop neighbors. We remove feature transformation and nonlinear activation that proved to be useless on collaborative filtering. Extensive experiments on three benchmark datasets show that our model outperforms various state-of-the-art models consistently.

\end{abstract}



\maketitle

\section{Introduction}

Personalized recommendation has been widely deployed in real-world applications, the core of which is capturing the users' preference accurately through their historical behavior. 
To address this problem, Collaborative Filtering (CF) ~\cite{su2009survey,koren2015advances} assumes that users with similar interaction history may have similar similar preference (e.g., click or watch) on items.
CF models have been widely studied and achieved great success in both industrial applications and academic research since their simplicity and effectiveness~\cite{koren2015advances}. 


The common paradigm of CF methods is to learn the representations (aka. embeddings) of users and items through user-item historical interactions. 
Early collaborative filtering methods, such as matrix factorization (MF), set the function of projecting the IDs of users and items as embedding individually.
In the view of graph, where the user's personal history can be viewed as a user-item interaction graph with user and item as nodes and user-item interactions as edges, early MF-based methods ignored the potential relationships existing in the user-item interaction graph. 
Later works find that further incorporating users' interaction history into user representations can improve the recommendation results~\cite{SVD,he2018nais}. 
These improvements demonstrate the benefits of modeling the one-hop neighbors of users in recommender systems.

The aforementioned recommendation models did not consider the implicit collaborative information of multi-hop neighbors. 
In recent years, inspired by the successes of Graph Neural Network (GNN) on graph-based tasks, employing GNN to model the multi-hop neighbors in user-item interactions have attracted a huge spike of interest in recommender systems.
For example, NGCF~\cite{NGCF19} performed Graph Convolution Network (GCN) in recommendation tasks and achieved the promising results. 
Each node of user-item interaction graph only has an ID as input which has no concrete semantics. 
LightGCN~\cite{he2020lightgcn} argued that the nonlinear transformation in \cite{NGCF19} is burdensome and proved it by experiments. DGCF~\cite{DGCF19} additionally combined GNN with users' potential intent and achieved the state-of-the-art results. 


However, previous works do not model the relational information in the neighborhood explicitly 
and ignore the implicit correlation among neighbors. The correlation of neighboring nodes implies the expression differences of different neighbors.
For example, in the user-item interaction graph, user's preferences can be represented by all neighboring items.
The user's embedding aggregates the information of its neighbor nodes during the training process, but it is difficult to avoid the information loss of its neighbor nodes, so we consider directly using the paired similarity between the user's neighbor nodes to obtain the attention coefficient. Through experiments, we found that in the recommendation scenario, calculating the attention coefficient of these neighbors directly through the paired similarity between the neighbors can achieve better results than the weight obtained by calculating the similarity between the user's embedding and its neighbor nodes.

To address the limitations mentioned above, we propose a novel Neighbor-Aware Graph Attention Network, termed \textbf{NGAT4Rec}, to capture the implicit correlations between nodes' neighborhood for recommendation task. Specifically, we introduce a neighbor-aware graph attention layer to compute the neighbor-aware attention coefficient for users’ neighboring items and items’ neighboring users. 
The neighbor-aware attention coefficient is employed in aggregation, which is a principal part of GNN, as aggregation coefficient. 
For a given user, the neighbor-aware attention coefficient for a neighboring item is computed by pairwise attention function including ReLU function and normalized dot product with all neighboring items of the given user. Because the embedding of user implicitly aggregates the embedding of it's neighboring items, we drop the attention between user and its neighboring items. The same applies to the case of a given item.
To simplify the calculation process, we compute the attention coefficient by averaging the results of each pairwise attention function among neighbors.
We use a linear combination of the neighbors' embeddings and attention coefficient to serve as next layer embeddings for every node. 
Having used the first-hop neighbors, we further stack more neighbor-aware graph attention layer to gather the influential signals from higher-order neighbors. Similar to LightGCN and DGCF, we remove feature transformation and nonlinear activation in our model as well. Extensive experiments on three public datasets verify the effectiveness of our method.
Recently, BGNN~\cite{BGNN} and NIA-GCN~\cite{NIA-GCN} are proposed to explicitly encode the neighboring node interactions pairwisely by element-wise product embedding for semi-supervised node classification task and recommendation task respectively. However, they do not study the impact of correlations among neighbors to aggregation coefficient
and still employ a nonlinear transformation which was proved to have negative effects on collaborative filtering \cite{he2020lightgcn}. Our works focus on assigning different importance to different neighbors of a given node by modeling the implicit correlations among these neighbors.

To summarize, the contributions of our work are as follows:
\begin{itemize}
    \item We emphasize the importance of implicit correlations among neighbors of each node on the user-item interaction graph and modeling of such correlations could lead to better representations in recommendation task.  
    \item We propose a novel model Neighbor-Aware Graph Attention Network (NGAT4Rec) for recommendation, which explicitly considers the implicit correlations among neighbors of each node. In addition, a max-M sub-graph sampling strategy is used to speed up the model training process.
    \item We conduct empirical studies on three million-size real-world datasets, demonstrating that the proposed method achieves competitive performances compared with the state-of-the-art.
\end{itemize}
\section{RELATED WORK}
In this section, we will briefly review several lines of works closely related to ours, including general recommendation and graph neural networks for recommendation.
\subsection{CF-Based Recommendation Models}
Recommendation systems typically use Collaborative Filtering (CF) to model users’ preferences based on their interaction histories~\cite{su2009survey,koren2015advances}. Among the various CF methods, item-based neighborhood methods~\cite{itemcf} estimate a user’s preference on an item via measuring its similarities with the items in her/his interaction history using a item-to-item similarity matrix. User-based neighborhood methods find similar users to the current user using a user-to-user similarity matrix, following by recommending the items in her/his similar users' interaction history. Matrix Factorization(MF)~\cite{MF4Rec} projects users and items into a shared vector space and estimate a user’s preference on an item by the similarity between user's and items' embedding vectors. 
BPR-MF~\cite{rendle2009bpr} optimizes the matrix factorization with implicit feedback using a pairwise ranking loss. 
However, above methods didn't consider relationship between user-item interaction pairs, so later works such as SVD++~\cite{SVD}, NAIS~\cite{he2018nais}, and ACF~\cite{ACF} treat historical interaction as features of users, and integrate the embeddings of historical items by average~\cite{SVD,he2018nais} or 
attention mechanism~\cite{ACF}. Autoencoders are also used in the recommendation task, such as Mult-VAE~\cite{MultVAE} and AutoRec~\cite{sedhain2015autorec}, to learn user preferences through the interaction history.
\subsection{Graph Neural Networks for Recommendation}
The main purpose of GNN models for recommendation is to integrate the distributed representations learning from user-item interaction graph.
GNN models can be divided into two categories: spatial GNN and spectral GNN~\cite{bruna2014spectral}. Spectral GNN, e.g. GCN~\cite{GCN}, is defined as performing convolution operations in the Fourier domain with spectral node representations. And Spatial GNN, e.g. GAT~\cite{GAT} and GraphSAGE~\cite{GraphSAGE}, instead performs convolution operations directly over the graph structure by aggregating the features from spatially close neighbors to a target node. 
To aggregate message from directly connected neighbors, different aggregators also account for multi-hop neighbors~\cite{atwood2016diffusion, GraphSAGE, xinyi2018capsule}. Moreover, non-linear aggregators are also employed in spatial GNNs such as max pooling~\cite{GraphSAGE}, capsule~\cite{verma2018graph}, and Long Short-Term Memory(LSTM)~\cite{GraphSAGE}. Furthermore, spatial GNN can be extended to graphs with complicated structure~\cite{park2019exploiting} and representations in hyperbolic space~\cite{chami2019hyperbolic}.
In most GNN models, the linear aggregators assume that neighbors are independent. BGNN~\cite{BGNN} considered importance of the interactions among neighbors, and used a bilinear aggregator to perform element-wise product between neighbors of every node.
GC-MC~\cite{GCMC} employs a graph convolution auto-encoder on user-item graph to solve the matrix completion task.
PinSage~\cite{PinSAGE} utilizes efficient random walks and graph convolutions to generate embeddings which incorporate both graph structure as well as node feature information. HOP-Rec~\cite{hoprec} employs label propagation and random walks on interaction graph to compute similarity scores for user-item pairs. NGCF~\cite{NGCF19} explicitly encodes the collaborative information of high-order relations by embedding propagation in user-item interaction graph.  Multi-GCCF~\cite{sun2019multigccf} constructs two separate user-user and item-item graphs. It employs a multi-graph encoding layer to integrate the information provided by the user-item, user-user and item-item graphs.
LightGCN~\cite{he2020lightgcn} proved that nonlinear feature transformation is useless in recommendation task that only uses the IDs of users and items as input. Distinct from mainstream CF models that parameterize user/item ID as a holistic representation only, DGCF~\cite{DGCF19} additionally separate the ID embeddings into K chunks, associating each chunk with a latent intent. Similar to \cite{BGNN}, NIA-GCN~\cite{NIA-GCN} deployed a pairwise neighborhood aggregation layer to capture relationships between pairs of neighbors.
Recently, many works applied graph neural networks to heterogeneous information networks(HIN)~\cite{HAN,HetGNN,HGT}. 
\citet{NIRec} proposed a novel formulation that captures the interactive patterns of HIN between each pair of nodes through their metapath-guided neighborhoods in recommendation tasks.

\section{Method}
\begin{figure*}[h]
  \centering
  \includegraphics[width=0.8\linewidth]{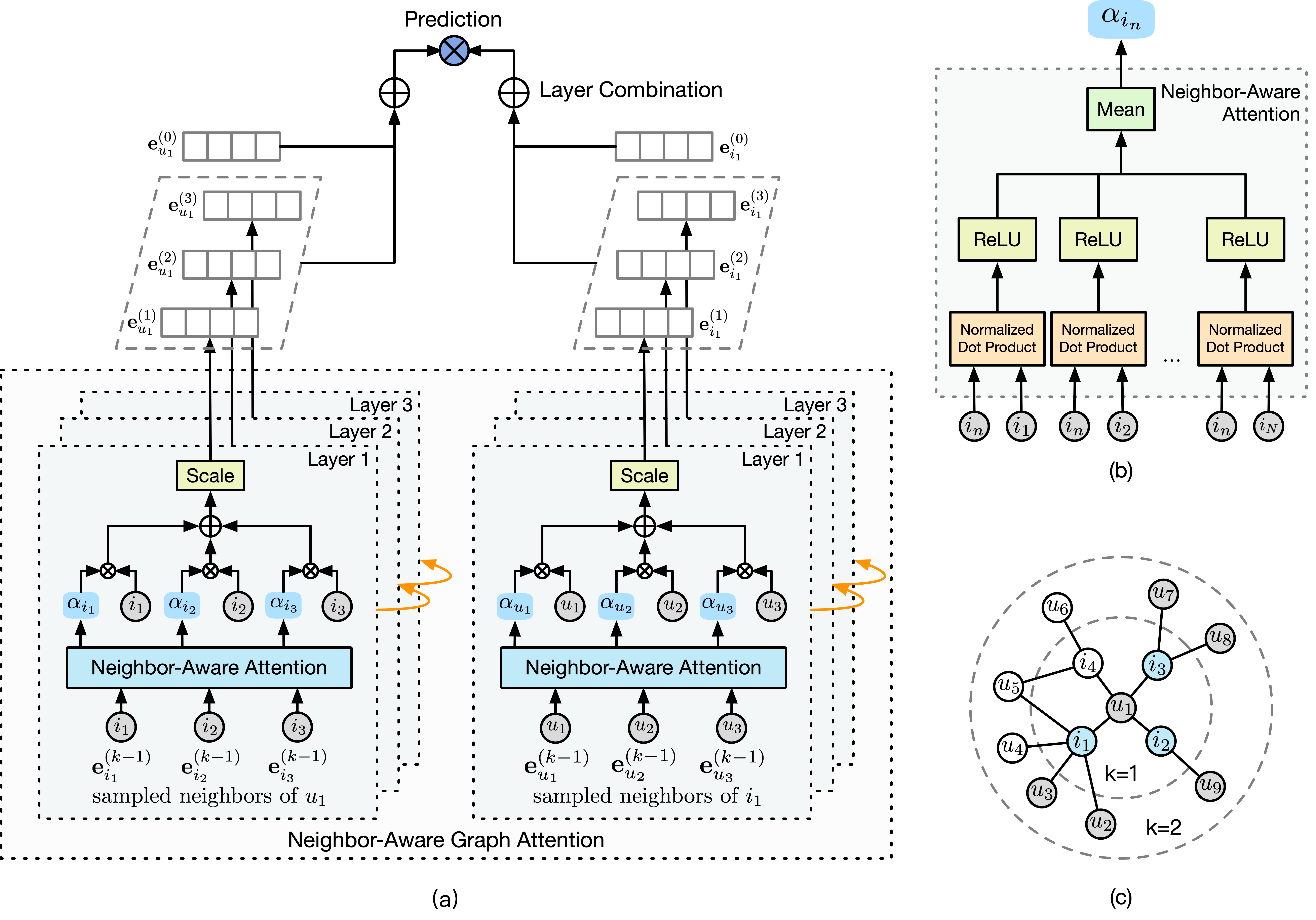}
  \caption{Neighbor-aware attention network model architecture and sampling strategy. (a) Neighbor-aware graph attention network; (b) Neighbor-aware graph attention mechanism; (c) Max-M sampling strategy, where k indicates the sample hop. We mark the sampled users as gray and the sampled items as blue.}
  \label{fig:NGAT} 
\end{figure*}
\begin{table}
  \caption{Notations and Explanations}
  \label{tab:note}
  \begin{adjustbox}{max width=\linewidth}
  \begin{tabular}{l l}
    \toprule
    Notation&Explanation\\
    \midrule
    $\mathcal{U}, \mathcal{I}$ & the set of users and items\\
    $\bm{e}_u^{(k-1)},\bm{e}_i^{(k-1)}$ & the embedding of user $u$ and item $i$ on $k$-th layer\\
    $\mathcal{N}_{u}$ & user $u$'s neighboring items in graph/sub-graph\\
    $\mathcal{N}_{i}$ & item $i$'s neighboring users in graph/sub-graph\\
    $\alpha_{i\rightarrow u}^{(k-1)}$ & the attention coefficient for $i$ aggregated to $u$ \\
    $|\mathcal{X}|$ & the number of elements in set $\mathcal{X}$\\
    $\Vert\bm{x}\Vert$ & the $\ell_2$ norm of vector/embedding $\bm{x}$\\
    $K$ & the total number of layers\\
    $M$ & upper bound of the number of sample neighbors\\
    $\Theta$ & the parameters of the model\\
    $\mathcal{B}$ & the sampled data of mini-batch\\
    $\mathcal{R}^{+}$ & the observed interactions between users and items\\
    $\mathcal{R}^{-}$ & the unobserved interactions between users and items\\
    $\sigma(\cdot)$ & nonlinear activation function\\
    $\lambda$ & the coefficient controlling $\ell_2$ regularization\\
  \bottomrule
\end{tabular}
\end{adjustbox}
\end{table}

Here, we introduce a new Neighbor-Aware Graph Attention Network for recommendation task, termed NGAT4Rec, which is illustrated in Figure \ref{fig:NGAT}(a). In this section, we will describe the model architecture in detail. We first introduce the embedding layer. Then we describe the neighbor-aware graph attention mechanism. Furthermore, we introduce a sample strategy to decrease the time and space complexity of neighbor-aware graph attention. Finally, we describe how we perform model prediction. The notations used in this paper are in Table 1.
 
\subsection{Embedding Layer}
There are two types of nodes in the graph abstracted from the recommendation scenario, namely user nodes and item nodes. In this paper, we use $u$ and $v$ as examples of user nodes, and $i$ and $j$ as examples of item nodes.
Embedding layer aims at mapping the ids of user $u$ and item $i$ into embedding vectors $\bm{e}_u^{(0)} \in \mathbb{R}^d$ and $\bm{e}_i^{(0)} \in \mathbb{R}^d$, where $d$ denotes the dimension of embedding. We use a trainable embedding lookup table to build our embedding layer for message propagation:
\begin{equation}
    \begin{split}
        \bm{E}_{\mathcal{U}} = [ \underbrace{\bm{e}_{1}^{(0)},\cdots,\bm{e}_{\left|\mathcal{U}\right|}^{(0)}}_{\text{users'}\,\text{embeddings}} ],\quad
        \bm{E}_{\mathcal{I}} = [ \underbrace{\bm{e}_{1}^{(0)},\cdots,\bm{e}_{\left|\mathcal{I}\right|}^{(0)}}_{\text{items'}\,\text{embeddings}} ]
    \end{split}
\end{equation}
where $\left|\mathcal{U}\right|$ is the number of users and $\left|\mathcal{I}\right|$ is the number of items. It should be noted that apart from the ID embedding of users and items, we did not introduce additional parameters.

\subsection{Neighbor-Aware Graph Attention Layer}

Graph attention mechanism in GAT~\cite{GAT} compute the propagation coefficient between the center node and its neighbor only by themselves. 
However, in recommendation tasks based on the user-item interaction graph, 
the importance of an item $i$ to a user $u$ can be represented by the correlations between $i$ and all neighboring items of $u$.
Therefore, as shown in Figure \ref{fig:NGAT}(b), we define a novel neighbor-aware graph attention mechanism that computes the neighbor-aware attention coefficient for users' items and items' users through implicit correlations on $k$-th layer:
\begin{equation}
\begin{split}
    \alpha_{i\rightarrow u}^{(k-1)} &= \operatorname{G}\left(\bm{e}_i^{(k-1)}, \{\bm{e}_j^{(k-1)} \bigm| j \in \mathcal{N}_u\}\right)\\
    &= g\left(\left\{f\left(\bm{e}_i^{(k-1)},\bm{e}_j^{(k-1)}\right) \Bigl|\, j \in \mathcal{N}_u\right\}\right) \\
    \alpha_{u\rightarrow i}^{(k-1)} &= \operatorname{G}\left(\bm{e}_u^{(k-1)}, \{\bm{e}_v^{(k-1)} \bigm| v \in \mathcal{N}_i\}\right)\\
    &= g\left(\left\{f\left(\bm{e}_u^{(k-1)},\bm{e}_v^{(k-1)}\right) \Bigl|\, v \in \mathcal{N}_i\right\}\right) \\
\end{split}
    \label{eq:neighbor_aware_attention}
\end{equation}
%
Where $\operatorname{G}(\cdot, \cdots)$ is the overall attention function. $f(\cdot, \cdot)$ is the pairwise attention function, and $g(\cdots)$ is the attention pooling function. Specially, we employ normalized dot product to calculate the pairwise attention. Furthermore, we perform $\operatorname{ReLU}$ function here to make sure the result of $f(\cdot, \cdot)$ is not negative. The pairwise attention function is defined as:
\begin{equation}
\begin{split}
    f\Bigl(\bm{e}_{i}^{(k-1)}, \bm{e}_{j}^{(k-1)}\Bigr) = \operatorname{ReLU}\left(\cos\Bigl(\bm{e}_{i}^{(k-1)}, \bm{e}_{j}^{(k-1)} \Bigr)\right) \\
    f\Bigl(\bm{e}_{u}^{(k-1)}, \bm{e}_{v}^{(k-1)}\Bigr) = \operatorname{ReLU}\left(\cos\Bigl(\bm{e}_{u}^{(k-1)}, \bm{e}_{v}^{(k-1)} \Bigr)\right)
\end{split}
\label{eq:3}
\end{equation}
In most previous works, attention function is defined by dot product~\cite{NLNN} or dot product with softmax~\cite{GAT}. However, normalized dot product can limit output from -1 to 1 in contrast to dot product may output very large or small values. Meanwhile, softmax is time-consuming to calculate the exponentials, therefore we utilize $\operatorname{ReLU}$ to obtain the same ability.

 

In order to simplify the calculation process, here we average the results calculated by each pairwise attention function. The final attention pooling function $g(\cdots)$ is defined as:
\begin{equation}
    \begin{split}
        \alpha_{i \rightarrow u}^{(k-1)} &= g\left(\left\{f\left(\bm{e}_i^{(k-1)},\bm{e}_j^{(k-1)}\right) \Bigl|\, j \in \mathcal{N}_u\right\}\right)\\
        &= \frac{1}{|\mathcal{N}_u|}\sum_{j\in \mathcal{N}_u}f\Bigl(\bm{e}_i^{(k-1)},\bm{e}_j^{(k-1)}\Bigr)\\
        \alpha_{u \rightarrow i}^{(k-1)} &= g\left(\left\{f\left(\bm{e}_u^{(k-1)},\bm{e}_v^{(k-1)}\right) \Bigl|\, v \in \mathcal{N}_i\right\}\right) \\
        &= \frac{1}{|\mathcal{N}_i|}\sum_{v\in \mathcal{N}_i}f\Bigl(\bm{e}_u^{(k-1)},\bm{e}_v^{(k-1)}\Bigr)
    \end{split}
    \label{eq:calculate_alpha}
\end{equation}
Where $|\mathcal{N}_u|$ and $|\mathcal{N}_i|$ are the size of user $u$'s and item $i$' neighbors in graph/sub-graph. It is very time-consuming to calculate the weights for all neighbors of each node in accordance with the above method. 
Therefore we employ a sampling strategy that introduced in Section~3.4.

 Through the Equation \ref{eq:calculate_alpha}, we can get the neighbor-aware attention coefficient of every node during message propagation. The message aggregation function on $k$-th layer is defined as:
\begin{equation}
\begin{split}
    \bm{e}_u^{(k)} &= \frac{1}{\sqrt{\left|\mathcal{N}_u\right|}}\sum_{i\in \mathcal{N}_u}\alpha_{i\rightarrow u}^{(k-1)} \cdot \bm{e}_i^{(k-1)} \\
    \bm{e}_i^{(k)} &= \frac{1}{\sqrt{\left|\mathcal{N}_i\right|}}\sum_{u\in \mathcal{N}_i}\alpha_{u\rightarrow i}^{(k-1)} \cdot \bm{e}_u^{(k-1)}
\end{split}
\label{eq:5}
\end{equation}
Considering that the norm of embedding after aggregation may be large, similar to \cite{he2020lightgcn}, we scaled these embedding by $\frac{1}{\sqrt{|\mathcal{N}_u}|}$ or $\frac{1}{\sqrt{|\mathcal{N}_i}|}$. 


\subsection{Layer Combination}
Having used the first-hop neighbors, we further stack more neighbor-aware graph attention layer to gather the influential signals from higher-order neighbors. For instance, the second-order connectivity like $u_1\rightarrow i_1 \rightarrow u_5$ suggests that $u_1$ and $u_5$ have a common interest when consuming $i_1$; meanwhile, the longer path $u_1\rightarrow i_1 \rightarrow u_5 \rightarrow i_4$ explores their interests via the collaborative signal. 
After $K$ layers, we sum the embeddings at different layers up as the final embeddings, as follows:
\begin{equation}
    \bm{e}_u^{*} = \sum_{k=0}^{K}\bm{e}_u^{(k)}, \quad
    \bm{e}_i^{*} = \sum_{k=0}^{K}\bm{e}_i^{(k)}
\end{equation}







\subsection{Sampling Strategy}
Previous works~\cite{he2020lightgcn, DGCF19} employed full neighborhood sets in message propagation, which is very time-consuming. There are often a large number of users and items in real recommendation scenarios. \cite{GraphSAGE,PinSAGE} sampled a fixed-size set of neighbors to perform inductive learning on graph. 
Suppose we sample $M$ neighbors for each node. If the number of neighbors of a node is less than M, there will be duplicate nodes in the neighbors sampled by \cite{GraphSAGE} for this node. These duplicate nodes are not necessary and import extra computation cost. Therefore, we performed max-M sub-graph sampling strategy, that is, in the sampled sub-graph, the number of neighbors of each node does not exceed M. For example, as shown in Figure \ref{fig:NGAT}(c), suppose $M=3$, $u_1$ and $i_1$ have 4 neighbors, and we randomly choose 3 neighbors $\{i_1, i_2, i_3\}$ and $\{u_1, u_2, u_3\}$, and neighbors of $i_2$ and $i_3$ are less than 3, therefore we choose all their neighbors.

\subsection{Model Learning}
\subsubsection{Model Prediction}
After embedding passing and aggregation with neighbor-aware attention mechanism, we obtained $\bm{e}_{u}^{*}$ and $\bm{e}_{i}^{*}$. In this way, we could predict the matching score between user and item by inner product:
\begin{equation}
    y_{ui} = {\bm{e}_u^{*}}^{\top} \bm{e}_i^{*}
\end{equation}

\subsubsection{Loss Function}
We employ Bayesian Personalized Ranking loss for optimization, which considers the relative order between observed and unobserved interactions. In order to improve the model's discrimination of similar positive and negative samples, we define BPR loss to optimize the model parameters $\Theta=\{\bm{e}_u^{(0)}, \bm{e}_i^{(0)} \mid u \in \mathcal{U}, i \in \mathcal{I}\}$:
\begin{equation}
\begin{aligned}
    \mathcal{L}_{\text{BPR}} = \frac{1}{|\mathcal{B}|}\sum_{(u,i,j)\in \mathcal{B}}-\sigma(y_{ui}-y_{uj}) + \lambda\|\Theta\|_{2}^{2}
\end{aligned}
\end{equation}
where $\mathcal{B} \subseteq \{(u,i,j)|(u,i)\in \mathcal{R}^{+}, (u,j)\in \mathcal{R}^{-}\}$ denotes the sampled data of mini-batch. $\mathcal{R}^{+}$ denotes observed interactions, and $\mathcal{R}^{-}$ is unobserved interactions. $\sigma$ is soft-plus function. $\lambda$ is the coefficient controlling $\ell_2$ regularization. 

\subsection{Time Complexity Analysis}
The time complexity of a single GAT attention head is $O(|V|FF^{'} + |E|F^{'})$, where $|V|$ is the number of nodes $(\#\mathit{users} + \#\mathit{items})$, $|E|$ is the number of edges; $F$ is the dimension of input features and $F'$ is the dimension of the output features. In LightGCN paper, authors remove feature transformation and nonlinear activation that proved to be useless on collaborative filtering, thus the time complexity of LightGCN on every layer is $O(|E|F^{'})$. According to Equation \ref{eq:3} and \ref{eq:calculate_alpha}, the time complexity of neighbor-aware attention mechanism is $O(K^2|V|F)$, where $K$ is the number of sampled neighbors. For message aggregation, we remove feature transformation and nonlinear activation as well, therefore according to Equation \ref{eq:5}, the overall time complexity of NGAT4Rec on every layer is $O(K^2|V|F)$. By setting the value of $K$ properly, the time complexity of NGAT4Rec can be on par with GAT.

\section{Experiments}

We first describe detailed experimental settings, then conduct performance comparison with state-of-the-art methods. We next conduct detailed comparison with LightGCN and DGCF. Further, to justify the designs in our model and reveal the reasons of its effectiveness, we perform hyper-parameter and ablation studies.
\subsection{Datasets \& Evaluation Metrics}
\begin{table}
\centering
\caption{Statistics of the datasets}
\label{tab:dataset}
\begin{tabular}{lrrrr}  
\toprule
 &Yelp2018&Amazon-Book&Kuaishou-Video\cr
\midrule
\#Users & 31,668 & 52,643 & 68,257\cr
\#Items & 38,048 & 91,599 & 165,397\cr
\#Interactions & 1.561m & 2.984m & 5.944m\cr
Density & 0.130\% & 0.062\% & 0.053\% \cr
\bottomrule
\end{tabular}

\end{table}
We evaluate the proposed model on three real-world representative datasets: Yelp2018,
Amazon-book
and Kuaishou-Video\footnote{Kuaishou is a famous Chinese short video APP. The dataset is available at https://github.com/ShortVideoRecommendation/Kuaishou-Video}. These datasets vary significantly in domains and sparsity. The statistics of the datasets are summarized in Table~\ref{tab:dataset}. To reduce the experiment workload and keep the comparison fair, we closely follow the training settings of the LightGCN and DGCF paper. We also use the same train/test splits in Yelp2018 and Amazon-Book that are provided in the GitHub repository of LightGCN\footnote{https://github.com/kuandeng/LightGCN} and DGCF\footnote{https://github.com/xiangwang1223/disentangled\_graph\_collaborative\_filtering}. Therefore the data split/setting is totally the same. 
 
For each dataset, the training set is constructed by $80\%$ of the historical interactions of each user, and the remaining as the test set. We randomly select $10\%$ of interactions as a validation set from the training set to tune hyper-parameters. We employ negative sampling strategy to produce one negative item that the user did not act before and treat observed user-item interaction as a positive instance. To ensure the quality of the datasets, we use the 10-core setting, i.e., retaining users and items with at least ten interactions. 

The evaluation metrics are recall@20 and NDCG@20 computed by the all-ranking protocol, i.e. all items that are not interacted by a user are regarded as the candidates, which are all same as the metrics used in LightGCN and DGCF paper.
\subsection{Baselines}
\begin{table*}
\centering
\setlength{\tabcolsep}{0.8em}
\caption{Overview performance comparison. Bold scores are the best and underlined scores are the second best.}
\begin{tabular}{lcccccccc}  
\toprule
&\multicolumn{2}{c}{Yelp2018}&\multicolumn{2}{c}{Amazon-Book}&\multicolumn{2}{c}{Kuaishou-Video} \\ \cmidrule(lr){2-3} \cmidrule(lr){4-5} \cmidrule(lr){6-7}
&Recall@20&NDCG@20&Recall@20&NDCG@20&Recall@20&NDCG@20\cr
\midrule
BPR-MF      & 0.0536 & 0.0423 & 0.0302 & 0.0224 & 0.0633 & 0.0648 \cr
GRMF        & 0.0561 & 0.0454 & 0.0352 & 0.0269 & 0.0742 & 0.0761 \cr
\midrule
PinSAGE     & 0.0516 & 0.0407 & 0.0296 & 0.0206 & 0.0641 & 0.0655 \cr
GAT         & 0.0543 & 0.0431 & 0.0326 & 0.0235 & 0.0682 & 0.0711 \cr
NGCF        & 0.0577 & 0.0476 & 0.0345 & 0.0264 & 0.0758 & 0.0773 \cr
NIA-GCN     & 0.0599 & 0.0491 & 0.0369 & 0.0287 & 0.0763 & 0.0779 \cr
LightGCN    & 0.0648 & 0.0528 & 0.0421 & \underline{0.0324} & 0.0796 & 0.0802 \cr
DGCF        & \underline{0.0654} & \underline{0.0534} & \underline{0.0422} & \underline{0.0324} & \underline{0.0801} & \underline{0.0808} \cr
NGAT4Rec & \textbf{0.0675} & \textbf{0.0554} & \textbf{0.0457} & \textbf{0.0358} & \textbf{0.0845} & \textbf{0.0864} \cr
\midrule
\%Improv.   & 3.21\% & 3.75\% & 8.29\% & 10.49\% & 5.49\% & 6.93\% \cr
\bottomrule
\end{tabular}
\label{PerformanceComparison}
\end{table*}
\begin{figure*}[]
  \centering
  \includegraphics[width=\linewidth]{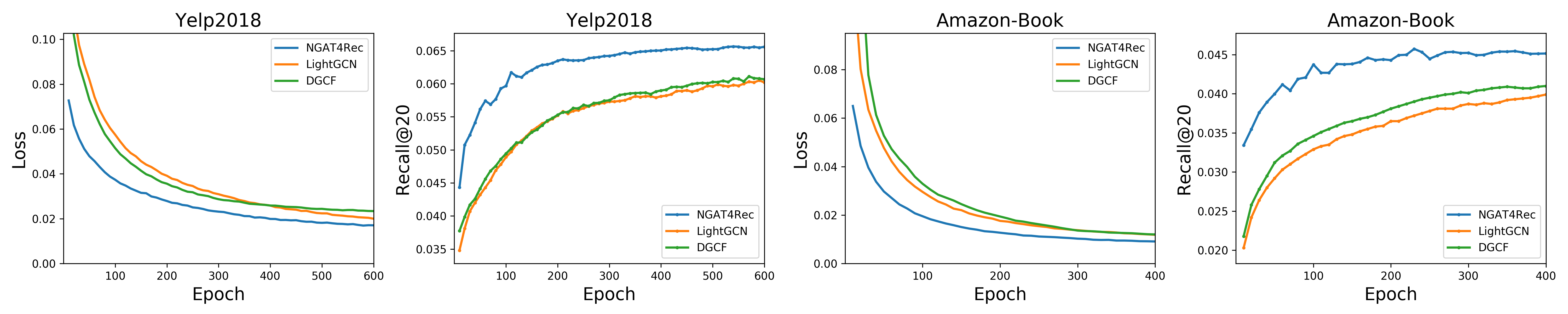}
  \caption{Training curves of NGAT4Rec, LightGCN and DGCF}
  \label{fig:Epoch}
\end{figure*}

To verify the effectiveness of our approach, we compare it  with the following baselines: 
\begin{itemize}
    \item \textbf{BPR-MF}~\cite{rendle2009bpr} optimizes the matrix factorization with implicit feedback using a pairwise ranking loss.
    \item \textbf{NGCF}~\cite{NGCF19} adopts three GNN layers on the user-item interaction graph, aiming to refine user and item representations via at most three-hop neighbors’ information.
    \item \textbf{PinSage}~\cite{PinSAGE} is designed to employ GraphSAGE~\cite{GraphSAGE} on item-item graph, that performs efficient, localized convolutions by sampling the neighborhood around a node and dynamically constructing a computation graph. In this work, we employ two graph convolution layers as suggested in \cite{PinSAGE} on user-item interaction graph.
    \item \textbf{GRMF}~\cite{GRMF} smooths matrix factorization by adding the graph Laplacian regularizer. For fair comparison on item recommendation, we change the rating prediction loss to BPR loss. 
    \item \textbf{GAT}~\cite{GAT} introduces a multi-head attention between node and neighbor nodes to calculate aggregate coefficients. We run the codes released by DGL~\cite{wang2019dgl}, and tuning the number of heads in $\left[2,4,8,16\right]$, and the feature dropout in $\left[0,0.1,0.2,0.5\right]$, and the negative slope in $\left[0,0.1,0.2\right]$.
    \item \textbf{NIA-GCN} ~\cite{NIA-GCN} modelings the interactions between neighbors with element-wise product by a bilinear neighborhood aggregator.  
    The Euclidean distance between users and items with their positive neighbors are added in loss function. The experimental setting are the same as the NIA-GCN paper.
    \item \textbf{LightGCN}~\cite{he2020lightgcn} removes the feature transformation and nonlinear activate function used in GCN. We implement the LightGCN by PyTorch and DGL~\cite{wang2019dgl}.
    \item \textbf{DGCF} ~\cite{DGCF19} deploys a graph disentangling module to separates the ID embeddings into several chunks, associating each chunk with a latent intent, and hires distance correlation as a regularizer to encourage independence of intents. 
\end{itemize}
All hyper-parameters and initialization strategies in baselines are either followed the suggestion from the methods’ authors or tuned on the validation sets. We report the results of each baseline under its optimal hyper-parameter settings.

\subsection{Hyper-parameter Settings}
Same as LightGCN and DGCF, the embedding size is fixed to 64 for all models and the embedding parameters are initialized with the Xavier method~\cite{XavierInit}. We optimize our model with Adam~\cite{kingma2014adam} and use the default learning rate of 0.0005 and default mini-batch size of 8192 (for 3 and 4 layers NGAT4Rec on Kuaishou-Video, we set the batch-size to 4096 to avoid out of memory error). For sampling from 1 hop to 4 hops, on Yelp2018 we set the maximum number of sampled neighbors to [120,120,120,120], and on Amazon-Book and Kuaishou-Video, in order to speed up training, we set the maximum number of sampled neighbors to [100, 80,60,40]. The $\ell_2$ regularization coefficient $\lambda$ is searched in the range of $\left\{{1e}^{-5}, {1e}^{-4}, {1e}^{-3}, {1e}^{-2}, 0.1, 0.5\right\}$, and in most cases the optimal value is ${1e}^{-4}$. The early stopping and validation strategies are the same as LightGCN. 
Our implementations are available in DGL~\cite{wang2019dgl} with PyTorch as backend\footnote{https://github.com/ShortVideoRecommendation/NGAT4Rec}. 

\subsection{Overall Performance Comparison}


Table \ref{PerformanceComparison} shows the performance comparison with our model and competing methods. And Figure \ref{fig:Epoch} shows the training curve of NGAT4Rec, DGCF and LightGCN, which are evaluated by training loss and testing recall per 10 epochs on Yelp2018 and Amazon-Book (results on Kuaishou-Video show exactly the same trend which are omitted for space). For a better comparison, we only retained BPR loss in the loss curves of the three models. Analyzing such performance comparison, we have the following observations:
\begin{table*}
\centering
\setlength{\tabcolsep}{0.8em}
\renewcommand{\arraystretch}{0.95}
\caption{Performance comparison of LightGCN, DGCF and our model $w.r.t$ number of layers. }
\begin{tabular}{clcccccccc}
\toprule 
\multirow{2}{*}{Layer \# } & \multirow{2}{*}{Method} & \multicolumn{2}{c} { Yelp2018 } & \multicolumn{2}{c} { Amazon-Book } & \multicolumn{2}{c} { Kuaishou-Video }\\
\cmidrule(lr){3-4} \cmidrule(lr){5-6} \cmidrule(lr){7-8}
 & & Recall@20 & NDCG@20 & Recall@20 & NDCG@20 & Recall@20 & NDCG@20 \cr
\midrule 
\multirow{4}{*} { 1 Layer } & LightGCN & 0.0577 & 0.0472 & 0.0376 & 0.0293 & 0.0695 & 0.0716\cr
& DGCF     & \textbf{0.0640} & \textbf{0.0522} & \underline{0.0399} & \underline{0.0308} & \underline{0.0712} & \underline{0.0757} \cr
& NGAT4Rec & \underline{0.0613} & \underline{0.0504} & \textbf{0.0405} & \textbf{0.0311} & \textbf{0.0749} & \textbf{0.0786}\cr
& \%Improv. & -4.22\% & -3.44\% & 1.5\% & 0.97\% & 5.19\% & 3.83\% \cr
\midrule 
\multirow{4}{*} { 2 Layers } & LightGCN & 0.0611 & 0.0504 & 0.0406 & 0.0312 & 0.0771 & 0.0780 \cr
& DGCF & \underline{0.0653} & \underline{0.0532} & \underline{0.0422} & \underline{0.0324} & \underline{0.0788} & \underline{0.0795} \cr
& NGAT4Rec & \textbf{0.0656} & \textbf{0.0540} & \textbf{0.0434} & \textbf{0.0339} & \textbf{0.0818} & \textbf{0.0839}\cr
& \%Improv. & 0.46\% & 1.50\% & 2.84\% & 4.63\% & 3.81\% & 5.93\% \cr
\midrule
\multirow{4}{*} { 3 Layers } & LightGCN & 0.0639 & 0.0525 & 0.0420 & 0.0322 & 0.0796 & 0.0802 \cr
& DGCF & \underline{0.0654} & \underline{0.0534} & \underline{0.0422} & \underline{0.0322} & \underline{0.0801} & \underline{0.0808} \cr
& NGAT4Rec & \textbf{0.0674} & \textbf{0.0554} & \textbf{0.0447} & \textbf{0.0350} & \textbf{0.0845} & \textbf{0.0864} \cr
& \%Improv. & 3.06\% & 3.75\% & 5.92\% & 8.70\% & 5.49\% & 6.93\% \cr
\midrule      
\multirow{4}{*} { 4 Layers } & LightGCN & \underline{0.0648} & \underline{0.0529} & \underline{0.0421} & \underline{0.0324} & \underline{0.0786} & \underline{0.0799} \cr
& DGCF & 0.0645 & 0.0521 & 0.0406 & 0.0312 & - & - \cr
& NGAT4Rec & \textbf{0.0669} & \textbf{0.0549} & \textbf{0.0457} & \textbf{0.0358} & \textbf{0.0812} & \textbf{0.0842} \cr
& \%Improv. & 3.28\% & 4.01\% & 8.62\% & 10.60\% & 3.33\% & 5.38\% \cr
\bottomrule
\end{tabular}
\label{LayerComparison}
\end{table*}
\begin{itemize}
    \item Our proposed NGAT4Rec consistently outperforms all baselines across three datasets. In particular, its relative improvements over the strongest baselines $w.r.t$ recall@20 are $3.21\%$, $8.29\%$ and $5.49\%$ on Yelp2018, Amazon-Book and Kuaishou-Video respectively and $w.r.t$ NDCG@20 are $3.75\%$, $10.49\%$ and $6.93\%$ on Yelp2018, Amazon-Book and Kuaishou-Video respectively. This demonstrates the effectiveness of NGAT4Rec. By neighbor-aware attention mechanism, NGAT4Rec is capable of exploring the implicit correlations among neighbors in an explicit way.
    It is worth noting that NGAT4Rec has achieved a significant improvement when using the sampling strategy.
    \item LightGCN and DGCF outperform all other baselines by a large margin which demonstrates 
    that nonlinear transformation is useless for the representation learning models of the recommendation system which has no other parameters except ID embedding of users and items. It is worth mentioning that the effect of NIA-GCN is worse than LightGCN and DGCF. The effect of BPR-MF is worse than that of GNN models, indicating that it is effective to modeling the multi-hop interactions between users and items. 
    \item Along the training process, NGAT4Rec consistently obtains lower training loss, which indicates that NGAT4Rec fits the training data better. Moreover, the lower training loss successfully transfers to better testing recall and NDCG, indicating the effectiveness of NGAT4Rec.
\end{itemize}

\subsection{Performance Comparison with LightGCN and DGCF}

We perform a detailed comparison between LightGCN, DGCF and our model. Table 4 shows the performance at different layers (from 1 to 4) and the percentage of relative improvement on each metric. The performance of DGCF on Yelp2018 and Amazon-Book from 1 layer to 3 layers are directly copied from the DGCF paper. The main observations are as follows:
\begin{figure}[]
  \centering
  \includegraphics[width=\linewidth]{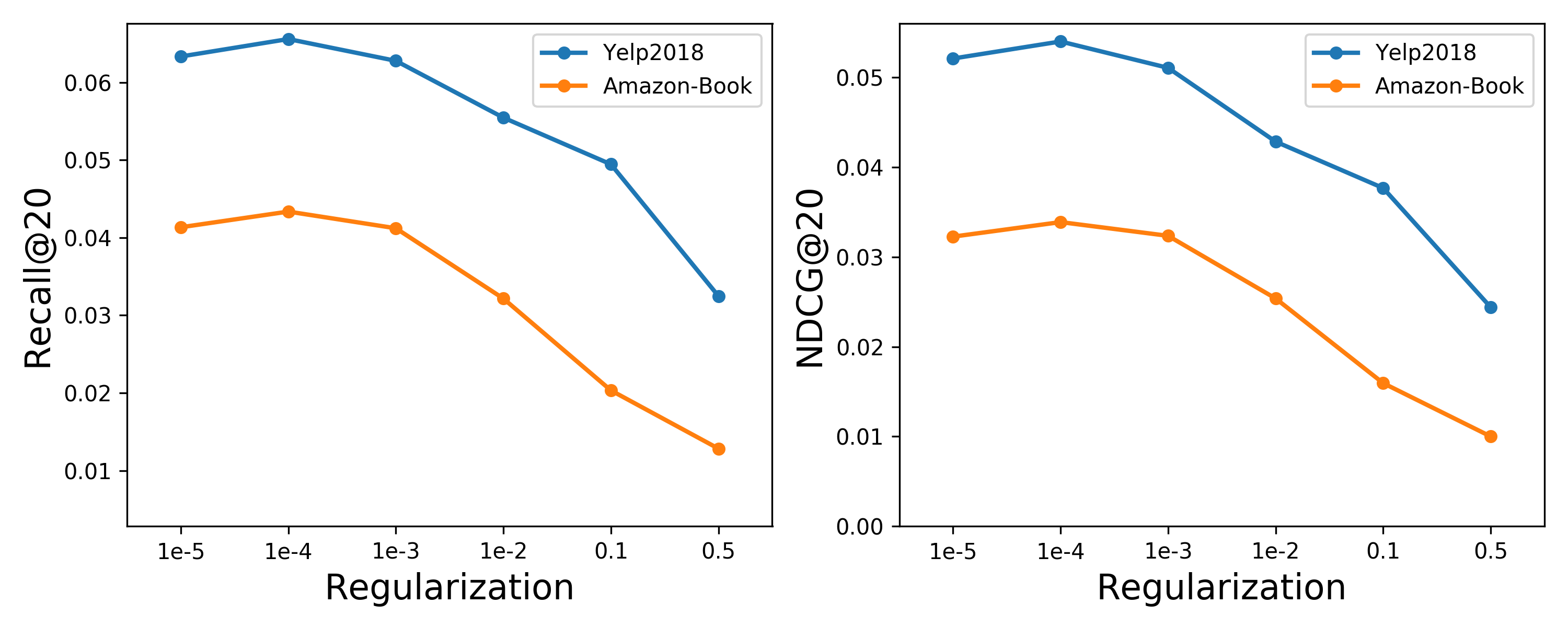}
  \caption{Performance of 2-layer NGAT4Rec w.r.t. different regularization coefficient $\lambda$.}
  \label{fig:Lambda}
\end{figure}
\begin{figure}[]
  \centering
  \includegraphics[width=\linewidth]{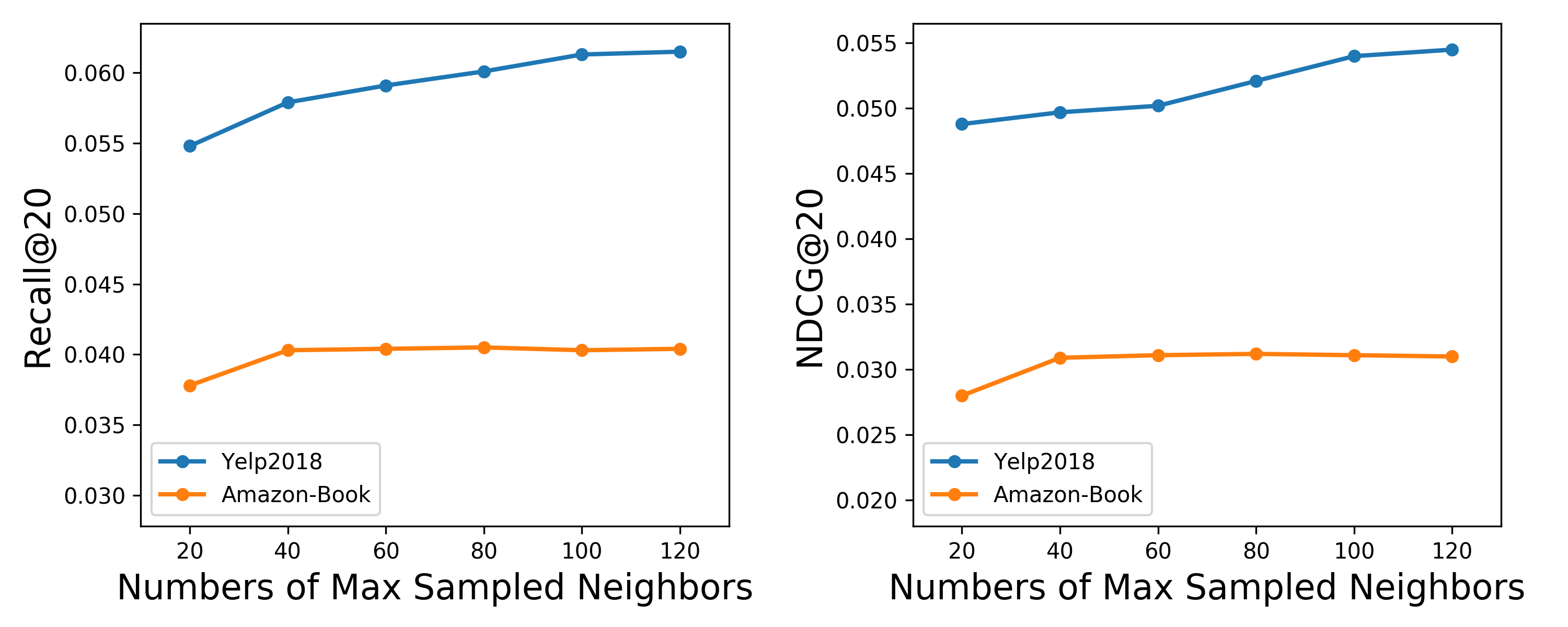}
  \caption{Performance of 1-layer NGAT4Rec w.r.t. different max number of sampled neighbors.}
  \label{fig:Neighbors}
\end{figure}

\begin{figure*}[]
  \centering
  \includegraphics[width=\linewidth]{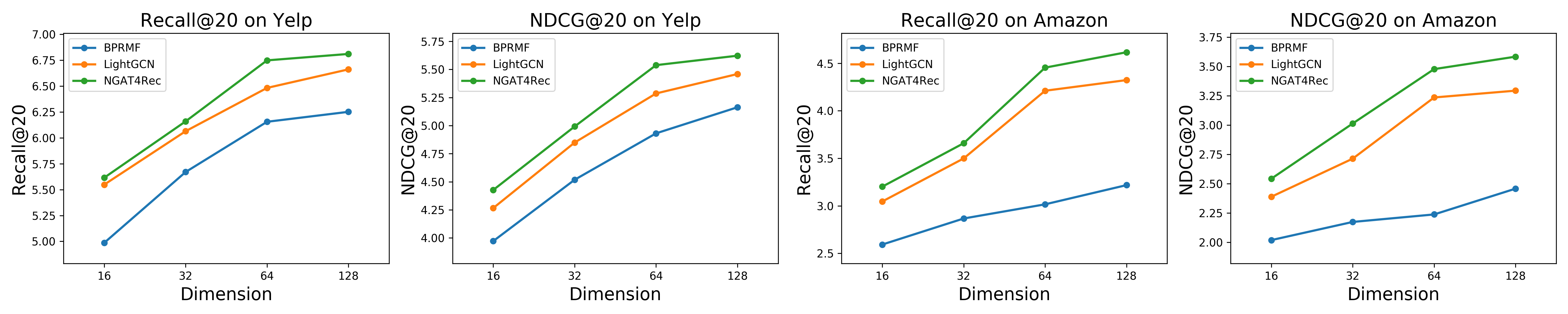}
  \caption{Effect curves w.r.t different dimension.}
  \label{fig:Dimension}
\end{figure*}
\begin{itemize}
    \item For our model, the increasing number of layers leads to better performance generally. The number of layers increased from 1 to 2 results in a largest performance gain, and our 3-layer model achieves the best results on Yelp2018 and Kuaishou-Video while 4-layer model achieves the best result on Amazon-book. 
    \item When the number of layers is 1, NGAT4Rec's effect on Yelp2018 is worse than DGCF, and on Amazon-Book and Kuaishou-Video, NGAT4Rec slightly ourperforms DGCF. When the number of layers increased from 2 to 3, NGAT4Rec outperforms DGCF. NGAT4Rec outperforms LightGCN on all layers across all three datasets. The experimental results show that compared to DGCF and LightGCN, NGAT4Rec can learn high-order information more effectively. When number of layers equals to 4, DGCF suffers from out of memory error on Kuaishou-Video.
    \item 2-layer NGAT4Rec performs better across three datasets than the best of LightGCN and DGCF, which means that when there is less information of multi-hop neighbors, our model can still achieve better results.
\end{itemize}

\subsection{Hyper-parameter Studies}


\subsubsection{Study on $\lambda$}

We investigate the performance change of 2-layer NGAT4Rec with different $\lambda$ on Yelp2018 and Amazon-Book. As shown in Figure \ref{fig:Lambda}, the performance of NGAT4Rec does not change significantly when $\lambda$ is not larger than $1\mathrm{e}{-3}$, and when $\lambda$ sets to $1\mathrm{e}{-4}$, NGAT4Rec achieves the best performance, therefore the optimal value for three datasets are $1\mathrm{e}{-4}$. 
When $\lambda$ is larger than $1\mathrm{e}{-3}$, the performance drops quickly, which indicates that too strong regularization will nagetively affect model training and is not encouraged.

\subsubsection{Study on Max Sampled Neighbors}

In this paper, we adopted the Max-M sampling strategy to train the model efficiently. For datasets with different sparsity, it is important for our model to select the appropriate number of sampled neighbors. Therefore we investigate the performance change of 1-layer NGAT4Rec w.r.t $M \in \left\{20,40,60,80,100,120\right\}$ on Yelp2018 and Amazon-Book. As shown in Figure \ref{fig:Neighbors}, on Yelp2018, the effect of NGAT4Rec increases with the increment of M, and the improvement doesn't obviously slow down after M is greater than 100, while on Amazon-Book, when M increases from 20 to 40, the effect of our model has been significantly improved, but when M is greater than 40, the effect of our model no longer changes. From Table \ref{tab:dataset}, we can know that Yelp2018 is denser than Amazon-Book and Kuaishou-Video, which means that each user and item has more neighbors. Therefore our experimental results show that the appropriate $M$ is closely related to the density of the dataset. On a sparser dataset, NGAT4Rec can achieve a significant improvement even with few sampled neighbors.

\subsubsection{Study on dimensionality $d$ of embeddings}

We conducted a dimension study on MF, LightGCN and NGAT4Rec on Yelp2018 and Amazon-Book. As shown in Figure \ref{fig:Dimension}, NGAT4Rec outperforms MF and LightGCN on all dimensions. As the dimension increases from 16 to 128, the performance of three models increases significantly. However, the improvemence slows down when the dimension increased from 64 to 128.

\subsection{Ablation and Effectiveness Analyses}
\begin{table}
\centering
\caption{Performance comparison for ablation study.}
\begin{adjustbox}{max width=\linewidth}
\begin{tabular}{lcccccc}  
\toprule
&\multicolumn{2}{c}{Yelp2018}&\multicolumn{2}{c}{Amazon-Book}&\multicolumn{2}{c}{Kuaishou-Video} \\ \cmidrule(lr){2-3} \cmidrule(lr){4-5} \cmidrule(lr){6-7}
&Recall@20&NDCG@20&Recall@20&NDCG@20&Recall@20&NDCG@20\cr
\midrule
$\text{NGAT4Rec}_\text{N}$ & 0.0564 & 0.0460 & 0.0380 & 0.0304 & 0.0755 & 0.0771 \cr
LightGAT-mlp           & 0.0603 & 0.0489 & 0.0396 & 0.0309 & 0.0793 & 0.0805 \cr
LightGAT-dp            & 0.0617 & 0.0504 & 0.0409 & 0.0314 & 0.0788 & 0.0800 \cr
NGAT4Rec               & 0.0675 & 0.0554 & 0.0457 & 0.0358 & 0.0845 & 0.0864 \cr
\midrule
$\%$vs $\text{NGAT4Rec}_\text{N}$ & 19.68\% & 20.43\% & 20.26\% & 17.76\% & 11.92\% & 12.06\% \cr
$\%$vs LightGAT-mlp & 11.94\% & 13.29\% & 15.40\% & 15.80\% & 6.56\% & 7.33\% \cr
$\%$vs LightGAT-dp  & 9.40\% & 9.92\% & 11.74\% & 14.01\% & 7.23\% & 8.0\% \cr
\bottomrule
\end{tabular}
\end{adjustbox}
\label{tab:ablation}
\end{table}
\subsubsection{Impact of Nonlinear Transformation}
To study the impact of nonlinear transformation, we implemented NGAT4Rec with nonlinear transformation noted as $\text{NGAT4Rec}_\text{N}$. In the $k$-th layer, we perform nonlinear transformations on the $\mathbf{e}_u^{(k-1)}$ and $\mathbf{e}_i^{(k-1)}$:
\begin{equation}
    \mathbf{p}_u^{(k-1)} = \sigma(\mathbf{W}_u^{(k-1)}\mathbf{e}_u^{(k-1)}),\quad
    \mathbf{p}_i^{(k-1)} = \sigma(\mathbf{W}_i^{(k-1)}\mathbf{e}_i^{(k-1)})
\end{equation}
Where $\sigma(\cdot)$ is the ReLU function. Then we use $\mathbf{p}_u^{(k-1)}$ and $\mathbf{p}_i^{(k-1)}$ to perform neighbor-aware attention, message propagation and layer combination. The hyperparameter settings of $\text{NGAT4Rec}_\text{N}$ on each dataset are the same as NGAT4Rec. Table \ref{tab:ablation} shows the performance comparison between $\text{NGAT4Rec}_\text{N}$ and NGAT4Rec. The improvements of NGAT4Rec over $\text{NGAT4Rec}_\text{N}$ $w.r.t$ recall@20 are $19.68\%$, $20.26\%$ and $11.92\%$ on Yelp2018, Amazon-Book and Kuaishou-Video respectively and $w.r.t$ NDCG@20 are $20.43\%$, $17.76\%$ and $12.06\%$ on three datasets respectively. The experimental results show that the nonlinear transformation is useless for the recommendation task that only takes the ID of user and item as input.

\subsubsection{Impact of Different Method to Compute Attention Coefficient}
In GAT~\cite{GAT}, the attention mechanism is a single-layer feedforward neural network, parametrized by a weight vector $\overrightarrow{\mathbf{a}}$ and applying the LeakyReLU function. In order to fairly compare the effect of the attention calculation method of GAT and NGAT4Rec in the recommendation task that only takes the ID of users and items as features, we implement two versions of GAT without nonlinear feature transformation, one applies MLP with softmax to compute the attention coefficient namely LightGAT-mlp:
\begin{equation}
    \alpha_{i\rightarrow u}^{(k-1)}=\operatorname{softmax}\left({\operatorname{LeakyReLU }\left(\overrightarrow{\mathbf{a}}^{\top}\left[\mathbf{W} \bm{e}_u^{(k-1)} \Bigl\|\, \mathbf{W}\bm{e}_i^{(k-1)}\right]\right)}\right)
\end{equation}
Where $d$ is the dimension of $\mathbf{e}_u^{(k-1)}$ and $\mathbf{e}_i^{(k-1)}$. $\mathbf{W} \in \mathbb{R}^{d \times d}$ is a weight matrix to transform the embedding, and $\overrightarrow{\mathbf{a}} \in \mathbb{R}^{2 d}$ is a weight vector. $\|$ denotes the concatenate operation.
The other uses the scaled dot product to calculate the attention on $k$-th layer noted as LightGAT-dp:
\begin{equation}
    \alpha_{i\rightarrow u}^{(k-1)} = \operatorname{softmax}\left(\frac{\mathbf{e}_u^{(k-1)^{\top}} \mathbf{e}_i^{(k-1)}}{\sqrt{d}}\right)
\end{equation}
Table \ref{tab:ablation} shows the performance comparison between LightGAT-mlp, LightGAT-dp and NGAT4Rec. The improvements of NGAT4Rec over LightGAT-mlp $w.r.t$ recall@20 are $11.94\%$, $15.40\%$ and $7.23\%$ on Yelp2018, Amazon-Book and Kuaishou-Video respectively and $w.r.t$ NDCG@20 are $13.29\%$, $15.80\%$ and $7.33\%$ on three datasets respectively. The improvements of NGAT4Rec over LightGAT-dp $w.r.t$ recall@20 are $9.4\%$, $11.74\%$ and $7.23\%$ on Yelp2018, Amazon-Book and Kuaishou-Video respectively and $w.r.t$ NDCG@20 are $9.92\%$, $14.01\%$ and $8.0\%$ on three datasets respectively. The experimental results indicate the effectiveness of the neighbor-aware attention mechanism on the user-item interaction graph.


\section{Conclusion}
In this paper, we proposed NGAT4Rec, a new GNN-based collaborative filtering model that captures the implicit correlations of neighbors. The key of NGAT4Rec is the novel neighbor-aware graph attention mechanism which assigns different importance to different neighbors of a given node by averaging the attention computed pairwisely among these neighbors. Extensive experiments show that the neighbor-aware graph attention layer utilized in NGAT4Rec is efficient. Our model leveraging neighbor-aware attention mechanism has achieved state-of-the-art performance across three real-world recommendation datasets. 

\balance
\bibliographystyle{ACM-Reference-Format}
\bibliography{sample-base}

\end{document}